\documentclass{JAC2003}

\usepackage{graphicx}
\usepackage{booktabs}

\newcommand\bgamma{\mbox{\boldmath$\gamma$}}
\newcommand\beq{\begin{equation}}
\newcommand\eeq{\end{equation}}
\newcommand\beqa{\begin{eqnarray}}
\newcommand\eeqa{\end{eqnarray}}

\begin{document}
\title{QCD ORIGIN OF STRONG MAGNETIC FIELD IN COMPACT STARS\thanks{Work partially supported by the 
Grant-in-Aid for the Global COE Program 
``The Next Generation of Physics, Spun from Universality and Emergence''
from the Ministry of Education, Culture, Sports, Science and Technology
(MEXT) of Japan  and the 
Grant-in-Aid for Scientific Research (C) (16540246, 20540267).}}

\author{T. Tatsumi\thanks{tatsumi@ruby.scphys.kyoto-u.ac.jp}, Department of Physics, Kyoto University, Kyoto 606-8502, Japan}

\maketitle

\begin{abstract}
 Some magnetic properties of quark matter and a microscopic origin of the strong magnetic field in compact stars are discussed; ferromagnetic order is discussed with the Fermi liquid theory and possible appearance of spin density wave is suggested within the NJL model. Implications of these magnetic properties are briefly discussed for compact stars.
\end{abstract}

\section{INTRODUCTION AND MOTIVATION}

Nowadays there have been many works about the QCD phase diagram. Here, we are concentrated in magnetic properties of quark matter and their implications on compact star phenomena. Let's begin with a simple question: what is and where can we expect magnetism in QCD. At low densities we may expect the appearance of pion condensation in hadronic matter, where the classical pion field develops , followed by the specific spin-isospin order of nucleons \cite{pio}. In quark matter we shall see a non-uniform phase (called dual chiral density wave (DCDW) phase), accompanying the restoration of chiral symmetry, where the pseudoscalar condensate $\langle {\bar q}i\gamma_5\tau_3 q\rangle\neq 0$ as well as the scalar condensate spatially oscillates \cite{nak}. Accordingly the magnetization of quark matter also oscillates like spin density wave (SDW) in condensed matter physics. Furthermore we may expect a ferromagnetic phase at some density region \cite{tat00}.

On the other hand such magnetic properties should have some implications on the compact star phenomena. In particular it has been well known that there is a strong magnetic field in compact stars. The origin of such strong field is not clear even now, and it has been a long-standing problem since the first discovery of pulsars in early seventies. The recent discovery of magnetars seems to revive the problem again \cite{mag}. Their magnetic field amounts to $O(10^{15}{\rm G})$ from the $P-{\dot P}$ diagram. At present many people believe the inheritance of the magnetic field from the progenitor main-sequence stars or dynamo scenario due to the electron current. We consider here a microscopic origin of magnetic field by examining a possibility of spontaneous spin polarization in quark matter.   

Next, we consider a possibility of a non-uniform state in the vicinity of the chiral transition. Recently there have done many works about the non-uniform state at moderate densities. 
We shall see that this phase exhibits an interesting magnetic property like SDW.

\section{FERROMAGNETIC TRANSITION}

The first study about the ferromagnetism in quark matter has been performed by using the Bloch idea about the ferromagnetism of itinerant electrons \cite{her66}. The mechanism is rather simple due to the Pauli principle: consider electrons interacting with each other by the Coulomb interaction in the background of the uniformly distributed positive charge to compensate the electron charge. Then the Fock exchange interaction gives a leading-order contribution. Then the electron pair with the same spin can effectively avoid the Coulomb interaction to give an attractive contribution to the total energy due to the Pauli principle. As the counter effect such polarized electron system costs more kinetic energy increase. So when the former effect becomes larger than the latter one, we can expect spontaneous spin polarization. A perturbative calculation of quark matter interacting with the one-gluon-exchange interaction shows the transition to ferromagnetic phase at the order of nuclear density. Applying this idea to a star with solar mass and radius of $10$km, we can roughly estimate the magnetic field of $O(10^{13-17}{\rm G})$. Thus we can feel that quark matter inside the core region may give an origin of magnetic field in compact stars. 

Recently we have studied the magnetic susceptibility $\chi_M$ to get more insight about the properties of ferromagnetic transition within the Fermi-liquid theory \cite{bay76}. $\chi_M$ is written in terms of the quasiparticle interactions,
\beq
\chi_M=\left(\frac{{\bar g}_D\mu_q}{2}\right)^2\frac{N(T)}{1+N(T)\bar
f^a},
\label{chim}
\eeq
where ${\bar g}_D\equiv \int_{|{\bf k}|=k_F} d \Omega_{{\bf k}}/4\pi g_D({\bf k})$ is the effective gyromagnetic ratio, $N(T)$ the effective density of states around the Fermi surface and $\bar f^a$ the spin dependent Landau-Migdal parameter \cite{tat082,tat083}. 
The spin susceptibility can be easily evaluated by using the OGE interaction. Then we can see that the Landau-Migdal (LM) parameters involve infrared (IR) divergences in gauge theories (QCD/QED). So we must take into account the screening effect at least 
to obtain the meaningful results. We have done it by calculating the quark polarization operator by the hard-dense-loop (HDL) resummation. As the results, we can see that the Debye screening for the longitudinal gluons surely improves the IR behavior, while the transverse gluons only receive the dynamic screening due to the Landau damping. 

\subsection{Zero temperature case}

There are still left the divergences in the LM parameters at $T=0$, but they cancel each other to give a finite $\chi_M$. 
Finally magnetic susceptibility is given as
a sum of the contributions of the bare interaction and the static 
screening effect,
\begin{eqnarray}
&&\left(\chi_M/ \chi_{\rm Pauli}\right)^{-1}_0=1
-\frac{C_f g^2\mu}{12\pi^2E_F^2k_F}\Big[m(2E_F+m)
-\nonumber\\
&&-\frac{1}{2}(E_F^2 + 4 m E_F - 2m^2)
\kappa\ln\frac{2}{\kappa} \Big],
\label{final}
\end{eqnarray}
with $\kappa=m_D^2/2k_F^2$ in terms of the Debye mass, $m_{D}^2\equiv g^2\mu k_{F}/2\pi^2$, and  $C_f=\frac{N_c^2-1}{2N_c}$. Thus the screening effect gives the $g^4\ln g^2$ term.

To demonstrate the screening effect, we show in Fig.~4.3 the magnetic
susceptibility. We assume a flavor-symmetric quark matter, 
$\rho_u=\rho_d=\rho_s=\rho_B/3$, and take the QCD coupling constant as
$\alpha_s\equiv g^2/4\pi=2.2$ and the strange quark mass $m_s=300$MeV inferred from the MIT bag model. 
\begin{figure}[h]
\begin{center}
\includegraphics[width=0.4\textwidth]{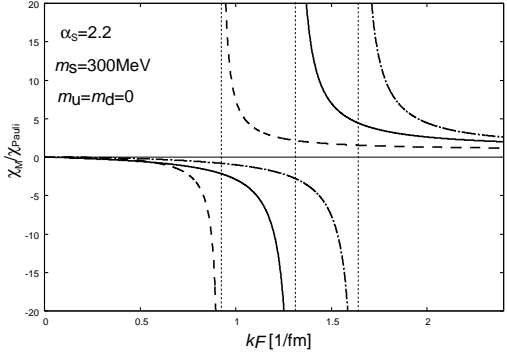}
\caption{Magnetic susceptibility at $T=0$. Screening effects are shown
 in comparison with the simple OGE case: the solid curve shows the result
 with the simple OGE without screening, while the dashed and dash-dotted ones 
 shows the screening effect with $N_f=1$ (only $s$ quark)and $N_f=2+1$
 ($u,d,s$ quarks), respectively.}
\end{center}
\label{result1}
\end{figure}
Note that the screening effect is qualitatively different, depending on the number of flavor $N_f$. The Debye mass is given by all the flavors,
\beq
m_D^2=\sum_{\rm flavors}\frac{g^2}{2\pi^2}k_{F,f}E_{F,f},
\eeq  
so that the $\kappa\ln (2/\kappa)$ term changes its sign for $\kappa=m_D^2/2k_F^2>2$. Thus we can see the screening {\it flavors} spontaneous magnetization in large $N_f$.


\subsection{Non-Fermi-liquid effect at finite temperature}

\begin{figure}[h]
\begin{center}
\includegraphics[width=0.4\textwidth]{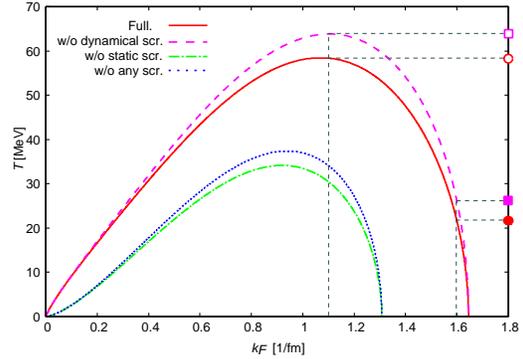}
\caption{Magnetic phase diagram in the density-temperature plane. The open (filled) circle indicates 
the Curie temperature at $k_F=1.1(1.6)$ fm$^{-1}$ while the squares show those without the $T^2 \ln T$ term.}
\label{Fig:diagram}
\end{center}
\end{figure}

We consider the low temperature case, $T/\mu\ll 1$, but usual low-temperature expansion can not be applied, since the quasiparticles exhibits an anomalous behavior near the Fermi surface. Since the longitudinal gluons are short ranged due to the Debye screening, their contributions are almost temperature independent. Thus the main contribution to the temperature dependence comes from the transverse gluons. Careful considerations about the quasiparticle energy show that quark matter behaves like {\it marginal Fermi liquid}, where the Fermi velocity 
and the renormalization factor vanish at the Fermi surface \cite{scha}. Such behavior is brought about by the transverse gluons. The magnetic susceptibility is then given as
\begin{eqnarray}
&&\left(\chi_M/\chi_{\rm Pauli}\right)^{-1} = 
\left(\chi_M/\chi_{\rm Pauli}\right)^{-1}_0\nonumber\\
&+&\frac{\pi^2}{6k_F^4} \left(2E_F^2-m^2+\frac{m^4}{E_F^2} \right)T^2
\nonumber\\
&+&\frac{C_fg^2u_F}{72}\frac{(2k_F^4+k_F^2m^2+m^4)}{k_F^4E_F^2} T^2\ln\left(\frac{\Lambda}{T}\right) \nonumber\\
&+&O(g^2T^2). 
\end{eqnarray}
with $u_F\equiv v_F/E_F$, where we can see the $T^2\ln T$ term appears as a novel non-Fermi-liquid effect, besides the usual $T^2$ term, It should be interesting to compare this term with other ones in specific heat or the superconducting gap energy. Furthermore, such logarithmic behavior also resembles the one by the spin fluctuations or paramagnons. Finally the phase diagram is presented in Fig.~\ref{Fig:diagram}, where we can also asses the importance of the non-Fermi-liquid effect. 

  Finally we present a phase diagram in Fig.~\ref{Fig:diagram}, where we can estimate the Curie temperature of several tens of MeV. We can also see how the non-Fermi-liquid effect works for the ferromagnetic transition.

\section{MAGNETISM AND CHIRAL SYMMETRY}

Recently there have been appeared many studies about the non-uniform states in QCD, stimulated by the development of the studies about the exact solutions in 1+1 dimensional models \cite{bas}. The formation of the non-uniform states in quark matter have been studied in relation to chiral transition \cite{nic}. The appearance of density waves or crystalline structures has been an interesting possibility at moderate densities. Note that the appearance of the non-uniform phase is not special, but rather familiar in condensed matter physics. In some cases it may exhibit an interesting magnetic property; the spin density wave (SDW) discussed by Overhauser is a typical example.  
\begin{figure}[h]
\begin{center}
\includegraphics[width=0.3\textwidth]{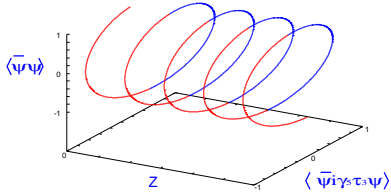}
\caption{Sketch of DCDW, where pseudoscalar density as well as scalar density oscillates along $z$ direction.}
\label{dcdw}
\end{center}
\end{figure}
In the previous paper \cite{nak} we have discussed the possibility of a density wave, where pseudoscalar density as well as scalar density oscillate in harmony along one direction, which is called dual chiral density wave (DCDW). 
\beqa
\langle \bar{\psi}\psi\rangle &=&\Delta \cos(\theta({\bf r})), \nonumber\\
\langle \bar{\psi}i\gamma_5 \tau_3\psi\rangle &=&\Delta\sin(\theta({\bf r})). 
\eeqa
The chiral angle $\theta({\bf r})$ is taken in the one dimensional form, $\theta({\bf r})={\bf q}\cdot{\bf r}$. The amplitude $\Delta$ generates the dynamical mass $M$, $M=-2G\Delta$, while $\theta$ produces the axial-vector field for quarks, $\tau_3\gamma_5\bgamma\cdot\nabla\theta/2=\tau_3\gamma_5\bgamma\cdot{\bf q}/2$.
The single-particle (positive) energy is then given by 
\beq
E^\pm_{\bf p}=[E_p^2+q^2/4\pm q\sqrt{p_z^2+M^2}]^{1/2},
\label{spect}
\eeq
with $E_p=\sqrt{p^2+M^2}$, depending on the spin degree of freedom. Accordingly the Fermi sea is split into two deformed ones: one is deformed in the prolate shape and the other in the oblate shape.

DCDW enjoys many interesting features. First, the symmetry breaking pattern is $T_{\hat p}\times U_{Q_5^3}(1)\rightarrow U_{{\hat p}+Q_5^3}$, which may be 1+1 dimensional analog of Skyrmion. Then the Nambu-Goldstone boson ("phason") has a hybrid nature of "pion" and "phonon". Secondly, a direct evaluation of the magnetization gives $\langle\sigma_{12}\rangle\propto \cos({\bf q}\cdot{\bf r})$, which means a kind of SDW.  In this case quark matter can be regarded as a kind of liquid crystal endowed with two-dimensional ferromagnetic order and one dimensional anti-ferromagnetic order. Note that magnetic field is globally vanished in this phase, but locally very strong.

\subsection{"Nesting" mechanism}

Here we discuss the mechanism for the formation of DCDW. There seems to be some confusions about it. In the references \cite{chiral} authors emphasized the nesting effect (or Overhauser effect) for the essential mechanism of chiral density waves, but there is little discussion about DCDW or other inhomogeneous phases. 
\begin{figure}[h]
\begin{center}
\includegraphics[width=0.4\textwidth]{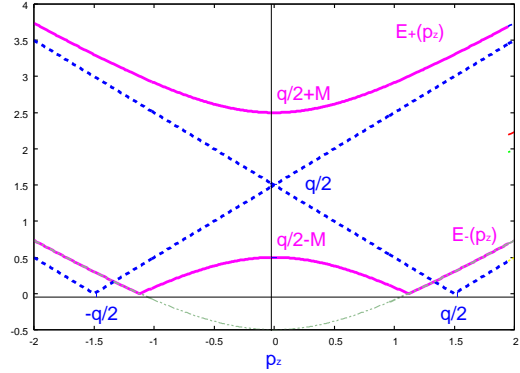}
\caption{Energy spectra for $p_\perp=0$ for $q/2>M$. Solid (magenta) curves show the one for massive quarks, while dashed (blue) curves for massless ones, $|p_z\pm q/2|$.}
\label{spectrum}
\end{center}
\end{figure}
For 1+1 dimensional case, we can immediately see that $q$ is given as $q=2\mu$ for given chemical potential $\mu$ \cite{bas}. This is simply because the energy spectrum (\ref{spect}) is reduced to $E_{\bf p}^{\pm}\rightarrow \left|\sqrt{p_z^2+M^2}\pm q/2\right|$ and $q$ is decoupled from $p_z$. Recall that the outstanding relation $q=2p_F$ is held in the usual density wave like CDW or SDW in the one dimensional system, due to the {\it nesting} effect of the Fermi surface. We can see that the similar mechanism works in the case of DCDW, but in somewhat different manner from the usual one. For the case, $M>q/2$, $E_{\bf p}^{\pm}$ are only shifted $\pm q/2$ from the free particle energy, so that formation of DCDW depends on the interaction strength like in the Stoner model. However, numerical calculation shows this is not the case: $q/2>M$ is always held in the DCDW phase. 
In Fig.~\ref{spectrum} we sketch the energy levels of the single quark energy for the case, $q/2>M$. Note that mass is generated by the interaction with DCDW in this case. So $E_{\bf p}^{\pm}$ can be regarded as a consequence of the switching of the interaction with DCDW between massless quarks with relative momentum $q$. For massless quarks, the two levels cross each other at $p_z=0$ for any $q$. Once the interaction with DCDW is present, mass is generated and two levels avoid the crossing with the energy gap, $2M$, at $p_z=0$ (magenta curves in Fig.~\ref{spectrum}). So if we choose $q=2\mu$ and fill the levels up to $p_F=\mu$, there is always the energy gain due to the interaction with DCDW. 
In the three dimensional case, the simple relation is no more held, but we can expect some reminiscence. Actually we can numerically check that the similar relation is held in the three dimensional case. In the vicinity of the critical end point we have seen that the chiral correlation function $\chi(q)$ diverges at finite $q$ of $q\sim 2p_F$, but the effective mass is almost vanished in this situation.

Thus we can understand DCDW with $q/2>M$ in terms of the "nesting" effect of the FErmi surface, while $q$ smoothly increases from zero for RKC. Their situation is very different from our case: the opposite relation, $|q/2<M|$, is realized in their case, and the level diagram looks very different from Fig.~(\ref{spectrum}). 

\subsection{Deformed DCDW}

Here we generalize the original DCDW by taking into account the symmetry breaking effect with the current quark mass $m_c\propto m_\pi^2$. 
Using a variational method, we can show that the functional form of the chiral angle in DCDW is deformed, satisfying the sine-Gordon (SG) equation, and thereby the allowed region of DCDW is extended. The stable solution is then given in terms of the Jacobian elliptic function with modulus $k$,
\beq
\theta=\pi+2{\rm am}\left(m_\pi^* z/k,k\right),
\label{elli}
\eeq
with the effective pion mass in medium, $m_\pi^*$. To recover the original DCDW in the chiral limit, we must require the following relation, 
\beq
qk=m_\pi^*\pi/{\bf K}
\eeq
with the complete elliptic integral of the first kind $\bf K$.
\begin{figure}[h]
\begin{center}
\includegraphics[width=0.35\textwidth]{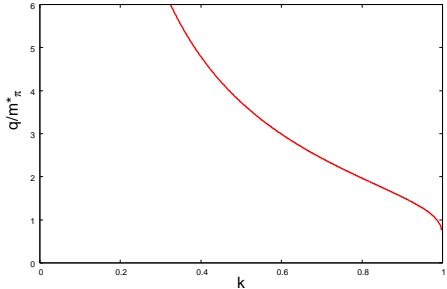}
\caption{Relation of the modulus $k$ and the parameter $q$. $k\rightarrow 0$ in the chiral limit.}
\label{wavenumber}
\end{center}
\end{figure}
Note that we can recover the SG equation in ref. \cite{sch} and $m_\pi^*\rightarrow m_\pi$ in the 1+1 dimensional case.

\section{CONCLUDING REMARKS}

We have discussed two kind of magnetic properties of quark matter separately, but a unified or comprehensive description of magnetism should be desired. Furthermore when we are interested in magnetism at moderate densities, we must take into account some non-perturbative effects explicitly. It may be one way to use the effective models of QCD, e.g. NJL model, to this end.     

Ferromagnetic order may have a direct implication on magnetic evolution of compact stars, but SDW order should also have some implications. DCDW may catalyse the elementary processes by providing extra momentum; for example it allows the quark $\beta$-decay process as neutrino emission during the thermal evolution. The magnetization is globally vanished there, but its fluctuation, $\langle {\bf M}^2\rangle$, becomes large. Accordingly the local strong magnetic field may induce new QED processes.

 For the present it needs more studies about the properties of the non-uniform states and relations among them. In particular, the comparison of DCDW and the real kink crystal in \cite{nic} is important, since they are typical structures in QCD, reflecting the different symmetries, $U(1)$ vs $Z_2$. 
More studies are needed including the symmetry breaking effect, thermal effect or model dependence.

The existing region of magnetism may be overlapped with color superconductivity.It is then interesting to elucidate the mutual relation in quark matter. 
Some works have been already done \cite{nak03}, but more studies are needed, including unconventional mechanism of pairing.

\end{document}